\documentclass[lettersize,journal]{IEEEtran}
\usepackage{amsmath,amsfonts}
\usepackage{algorithmic}
\usepackage{algorithm}
\usepackage{array}
\usepackage[caption=false,font=normalsize,labelfont=sf,textfont=sf]{subfig}
\usepackage{textcomp}
\usepackage{stfloats}
\usepackage{url}
\usepackage{verbatim}
\usepackage{graphicx}
\usepackage{cite}

\begin{document}

\title{Representations of Noisy $N$-Ports}

\author{Martin Bucher and Daniel Molnar\\
\today }

\markboth{Journal of \LaTeX\ Class Files,~Vol.~14, No.~8, August~2021}%
{Shell \MakeLowercase{\textit{et al.}}: A Sample Article Using IEEEtran.cls for IEEE Journals}


\maketitle

\begin{abstract}
Much has been written about the representation of noisy linear 2-ports. Here we present a theory
of noisy $N$-ports. We show how in the general case there are $(2N)!/(N!)^2$ equivalent representations
and give the transformations relating them. We also discuss singular cases in which some of 
the transformations are not possible as well as how to measure the noise properties of an $N$-port.
\end{abstract}


\section{Introduction}

\begin{figure}[!h]
\centering
\includegraphics[width=2.5in]{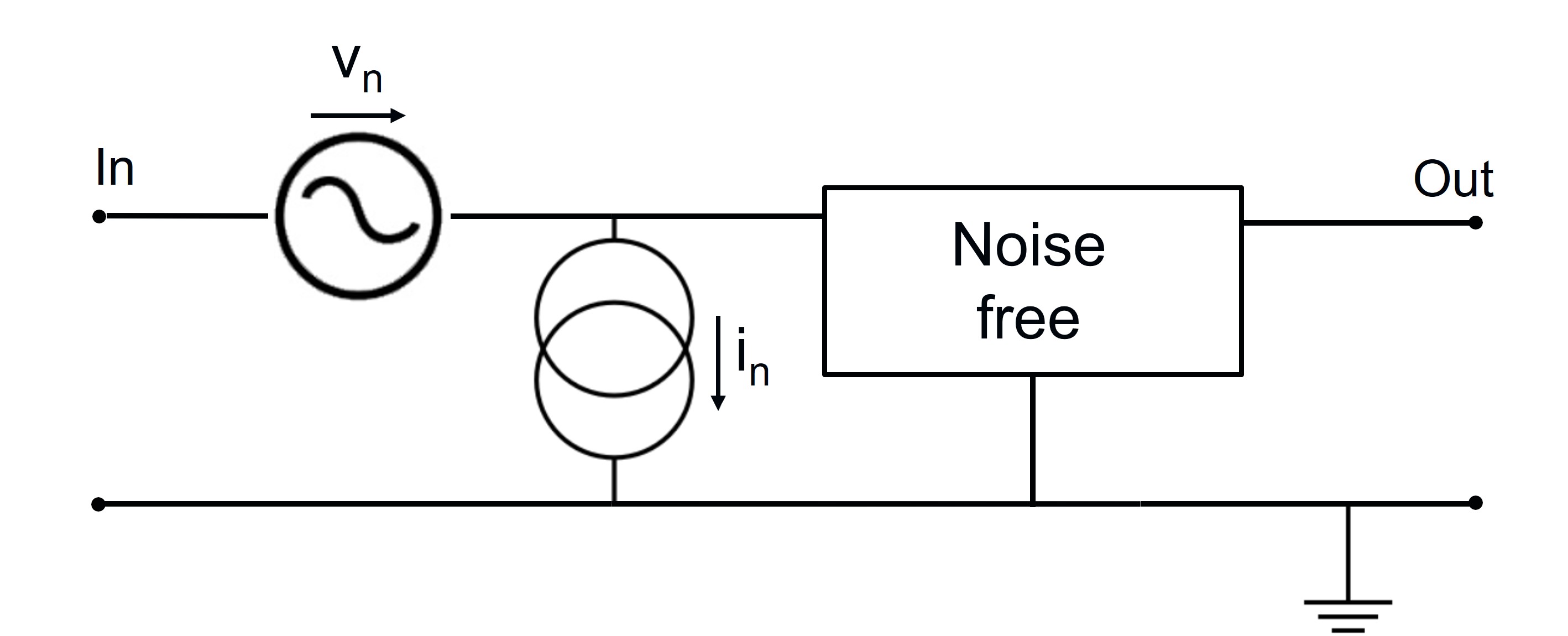}
\caption{Noisy Amplifier or 2-Port.}
\label{fig_zero}
\end{figure}

There is an extensive literature on noisy linear 2-ports. It is for example well known that
a noisy amplifier can be represented by an equivalent circuit consisting of a noiseless amplifier
with a voltage and current source placed on the input side as sketched in 
Fig.~\ref{fig_zero}.
(See for example \cite{reprNoisyTwoPorts,rfdesignNote}.)
As sketched, the circuit looks like an inhomogeneous 2-port, but  here 
the complex source amplitudes
$v_n$ and $i_n$ (at a particular angular frequency $\omega $) are taken to be determined stochastically
according to a Gaussian random ensemble. The correlation functions 
$\left\langle {v_n}^* v_n \right\rangle ,$
$\left\langle {i_n}^* i_n \right\rangle ,$
and 
$\left\langle {v_n}^* i_n \right\rangle ,$
which are described by four real parameters, 
completely suffice to determine the noise properties of the amplifier, or noisy
2-port.\footnote{Here we are assuming that the noise is completely characterized 
by its second-order correlation
functions, which is the case for Gaussian noise, for which 
there are no phase correlations
between different Fourier coefficients. For non-Gaussian noise, of which burst noise is one example, 
higher-order correlation functions as well are required to characterize the noise. However 
even when the noise is non-Gaussian, the two-point correlation functions considered here
provide a partial characterization and for many applications suffice.}
The two auto-correlations are real and positive, and the cross-correlation may be expressed as 
$$
\left\langle {v_n}^* i_n \right\rangle =(\alpha _n+j\beta _n)
\sqrt{
\left\langle {v_n}^* v_n \right\rangle 
\left\langle {i_n}^* i_n \right\rangle 
}
$$
where 
${\alpha _n}^2+
{\beta _n}^2\le 1.$

Noisy $N$-ports (where $N>2$) have been previously studied,
in particular in the work of Haus and Adler 
\cite{hausAndAdlerCanonical, optNoiseLinAmplifiers, hausAdlerMonograph}
and of Bosma \cite{bosmaThesis}. There the emphasis, generalizing
on earlier work by Mason \cite{mason}
on 
2-ports, was on studying the invariant properties of $N$-ports independent of
their embedding using lossless components to map the $N$ ports
into $N$ outer 
ports of the embedding network. Such an embedding network among other things
can transform impedances of ports and introduce feedback, both positive and 
negative.
Mason introduced an invariant know as the 
`unilateral power gain' that does not change as a result of the choice 
lossless 
embedding network. This line of investigation clarified the relation between feedback
and noise (or noise figure). 

Here our motivation is slightly different. We investigate how noise sources can be moved 
around to give a multitude of equivalent representations of a noisy
$N$-port. Our
underlying application is the modelling of global 21 cm experiments in radio astronomy, 
where the object is to make a absolute noise measurement of the sky signal 
at an accuracy to our knowledge higher than that required for other applications
(see  \cite{edgesDetectionPaper}, \cite{reachNaturePaper} and references therein).
The techniques presented here allow circuits with noisy $N$-ports and loops to be  
simplified by moving noise sources around. 
Related work includes work by Twiss generalizing Nyquist's theorem 
to lossy multiport devices
at a common temperature \cite{twiss} and work on noise 
matching for active phased array 
antennas \cite{warnickJensen,warnickJeffs,ivashina}. 

\begin{figure}[!t]
\centering
\includegraphics[width=2in]{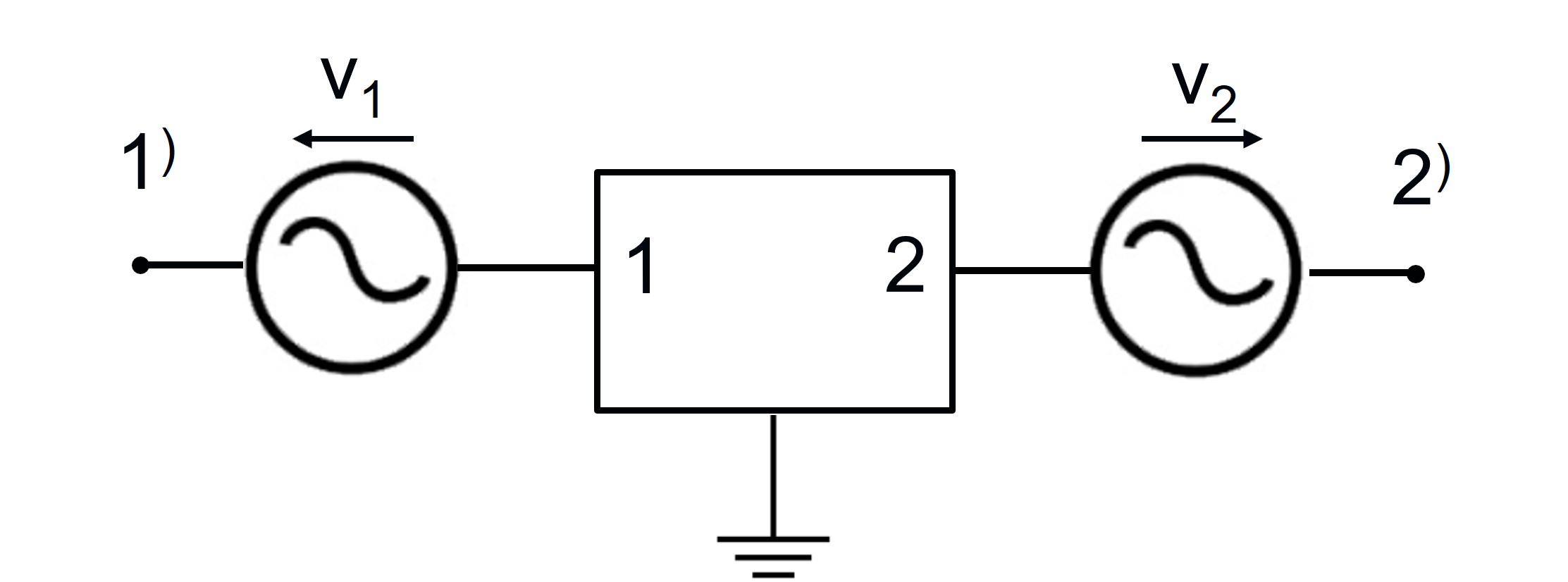}
\includegraphics[width=2in]{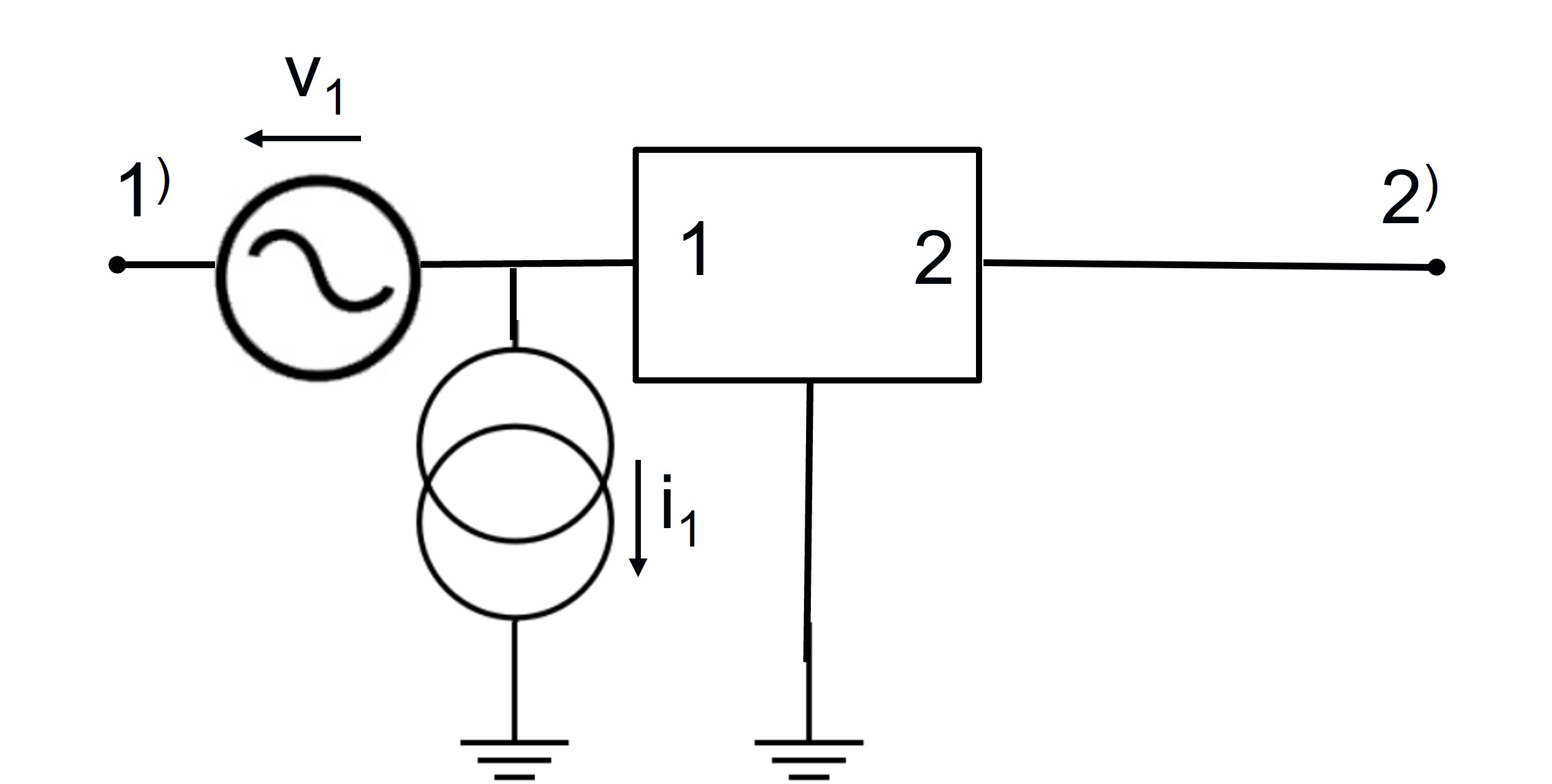}
\includegraphics[width=2in]{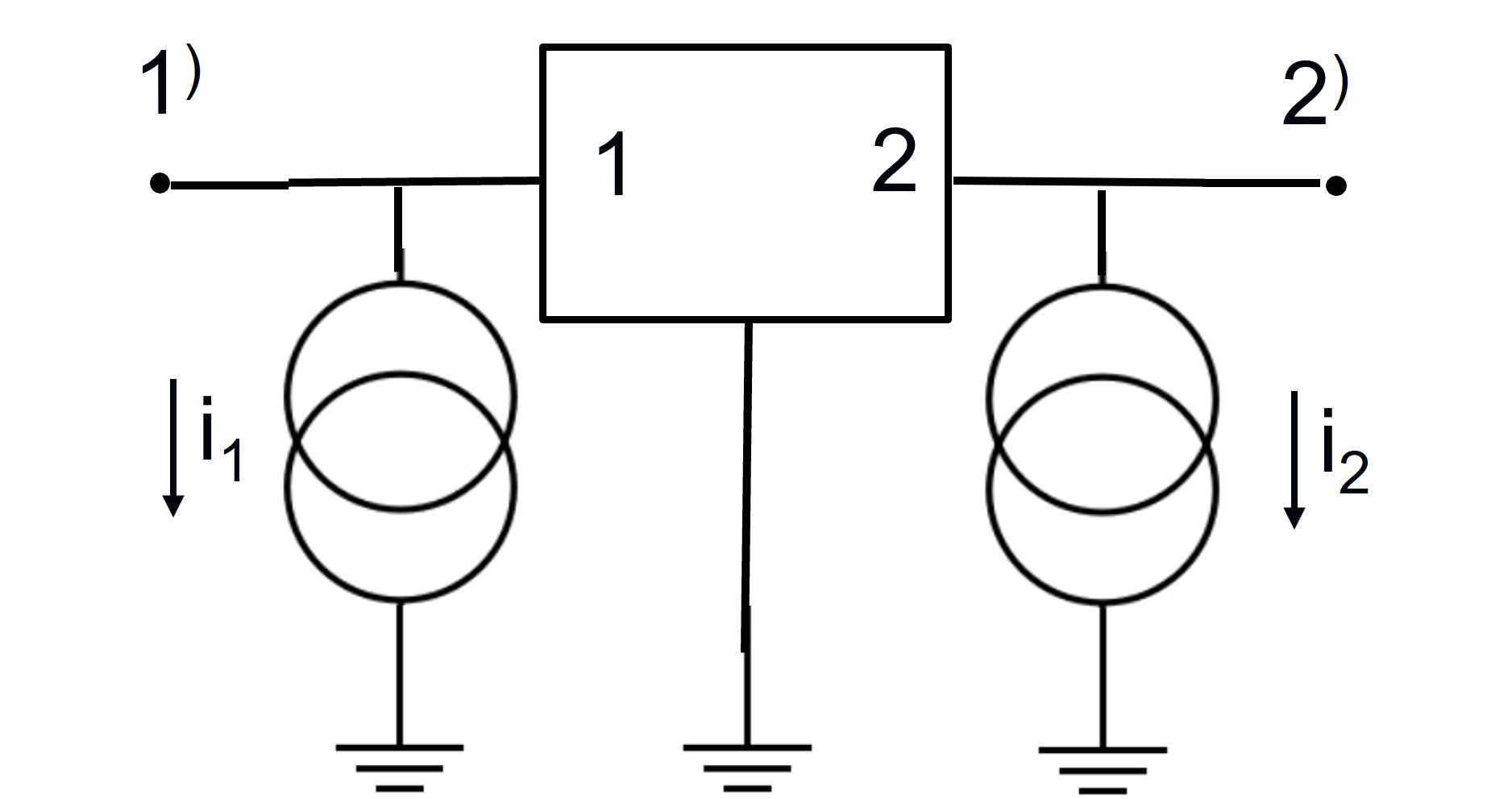}
\includegraphics[width=2in]{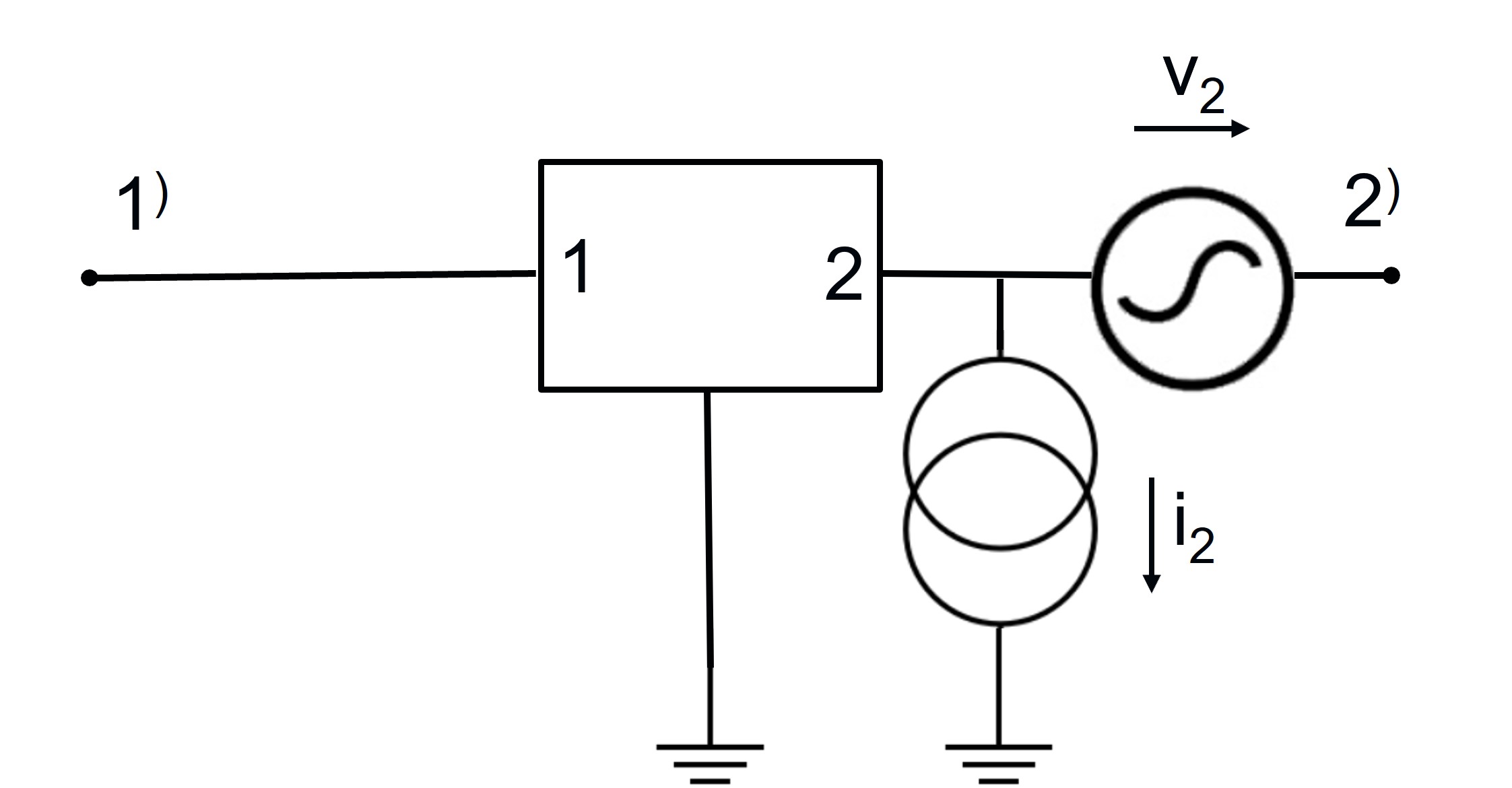}
\includegraphics[width=2in]{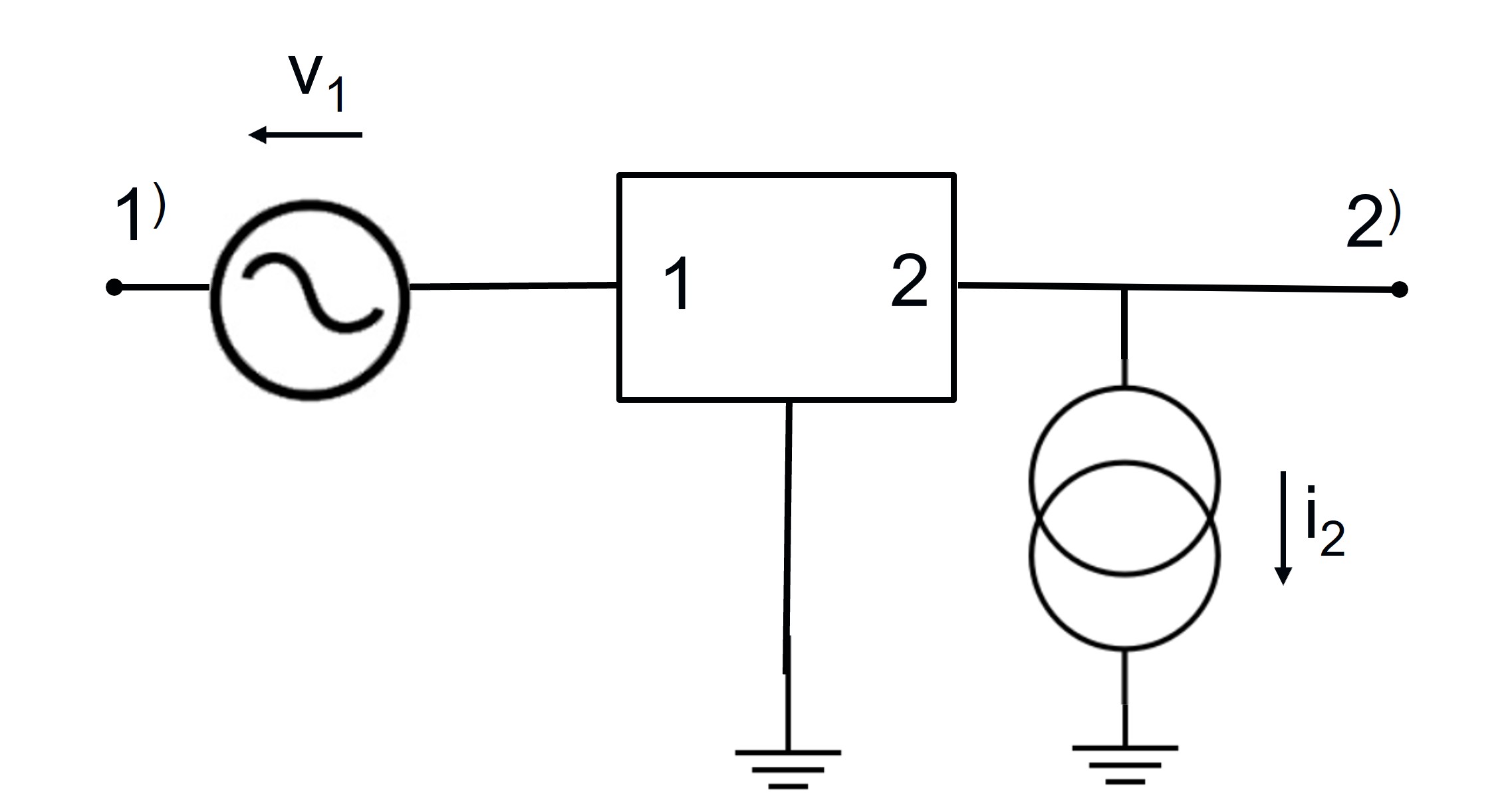}
\includegraphics[width=2in]{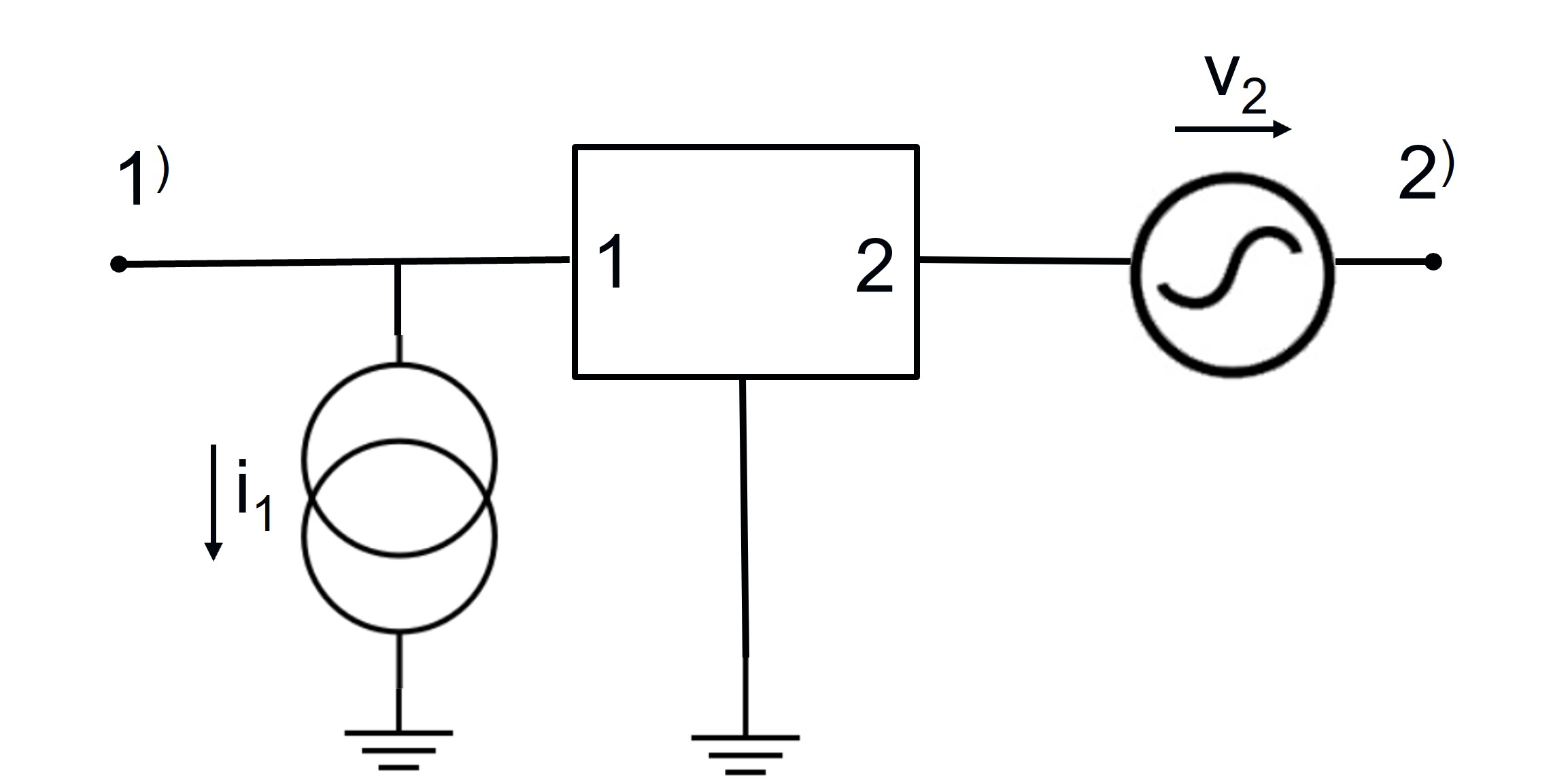}
\caption{Six Equivalent Noisy 2-Port Representations.}
\label{fig_zeroPlus}
\end{figure}

For the setup in Fig.~\ref{fig_zero}, when the amplifier is not unilateral, 
we can displace the noise sources to the output,
and in the general case there are six equivalent representations for the noisy 2-port,
as sketched in Fig.~\ref{fig_zeroPlus}. 

\begin{figure}[!h]
\centering
\includegraphics[width=2.5in]{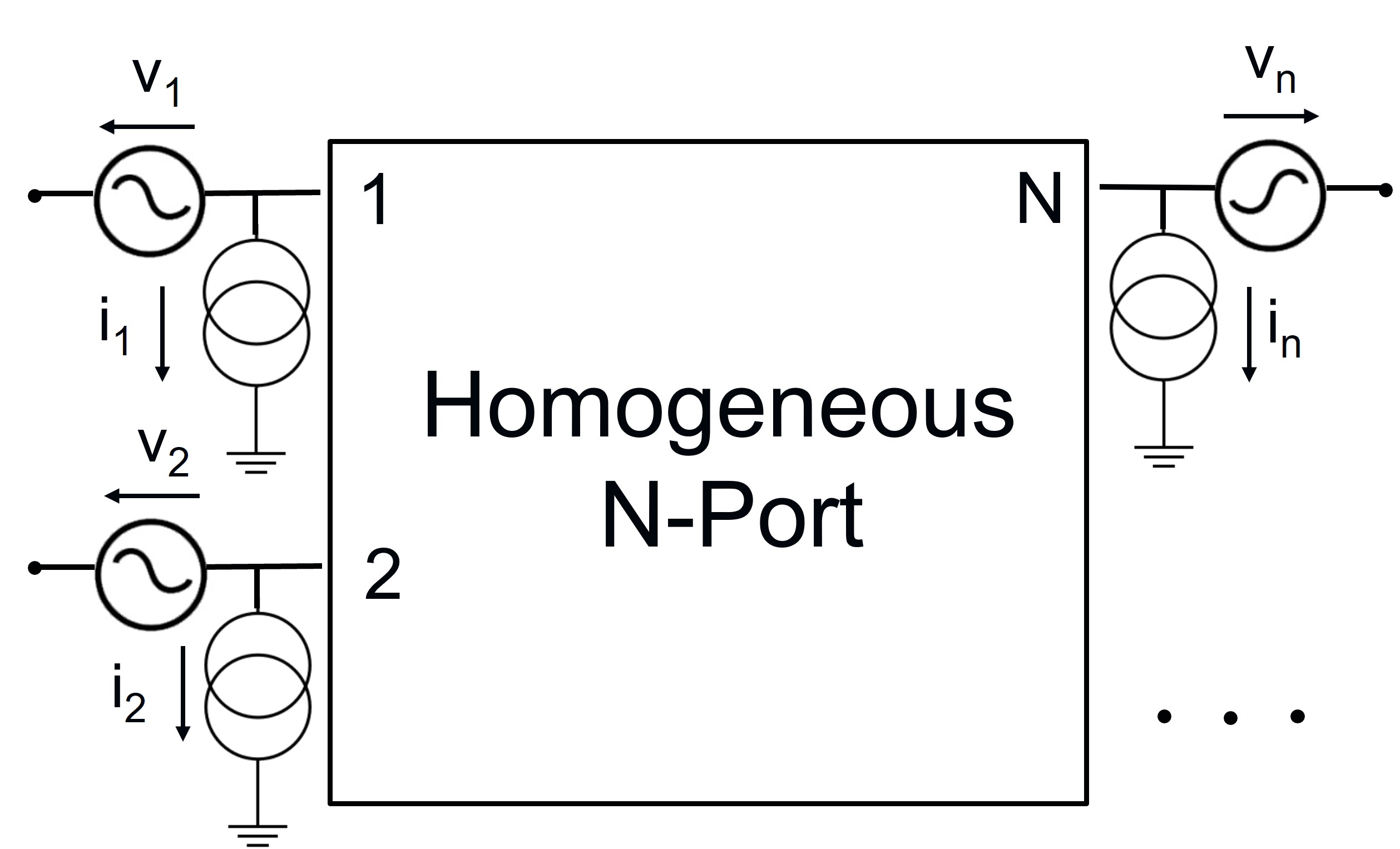}
\caption{General Noisy $N$-Port Representations.}
\label{fig_one}
\end{figure}

\begin{figure}[!h]
\centering
\includegraphics[width=2.5in]{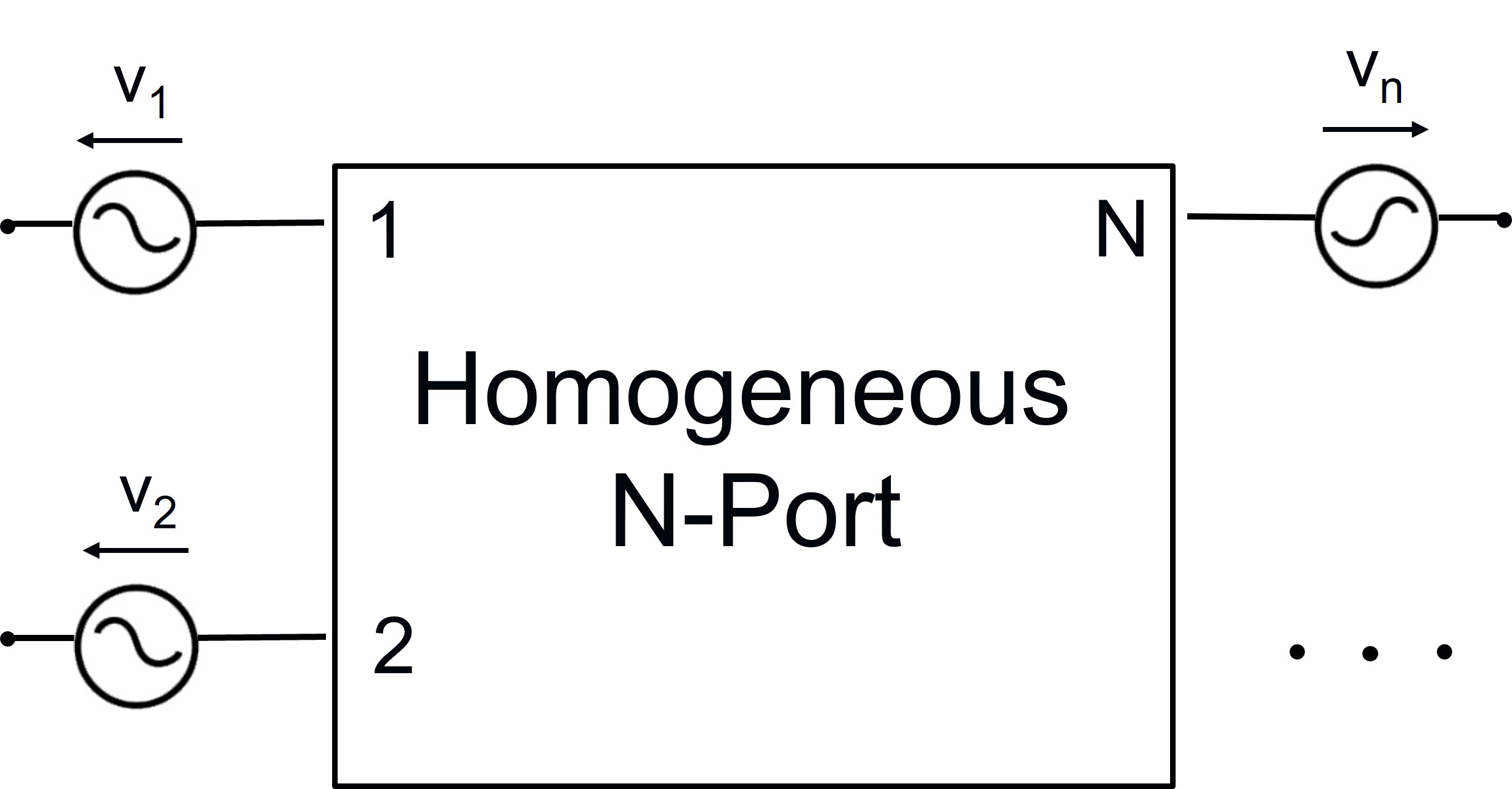}
\caption{Noisy $N$-Port Representation With Only Voltage Sources.}
\label{fig_two}
\end{figure}

We now proceed to generalize by describing a noisy $N$-port. Let 
$\bar v_I,$ $\bar \imath _I$ $(I=1..N)$ be the voltages and currents as added to the embedded 
homogeneous $N$-port, as illustrated in Fig.~\ref{fig_one}.
Similarly 
$v_I,$ $i _I$ 
are the voltages and currents as seen at the ports of the inhomogeneous $N$-port
and 
$v_I',$ $i _I'$ are the voltages and currents at the terminals of the embedded homogeneous
$N$-port, which may for example be related by the impedance matrix $Z_{IJ},$ so that 
\begin{equation} 
v_I'=Z_{IJ}i_J'
\label{eqn:ZmatHomo}
\end{equation} 
when this representation is not singular. When this homogeneous $N$-port is dressed with 
voltage and current sources as illustrated in Fig.~\ref{fig_one}, eqn.~(\ref{eqn:ZmatHomo})
is modified to become   
\begin{equation} 
(v_I -\bar v_I)
=Z_{IJ}
(i_J -\bar \imath _J).
\label{eqn:ZmatInhomo}
\end{equation} 
Here there are $2N$ voltage and current sources, but in the general case the number of sources 
can be reduced to $N.$
For example, when the Z representation of the embedded homogeneous N-port exists, by setting
\begin{equation} 
\bar{\bar{v}}_I=
\bar{v}_I-Z_{IJ}\bar \imath _J
\end{equation} 
we may eliminate all the $N$ current sources and obtain an equivalent representation with $N$ voltage 
sources, as sketched in Fig.~\ref{fig_two}. From the point of view of the external ports, from which
only $v_I,$ $i_I$ are visible, the representations in Figs.~\ref{fig_one} and \ref{fig_two} are 
indistinguishable. 

\begin{figure}[!h]
\centering
\includegraphics[width=2in]{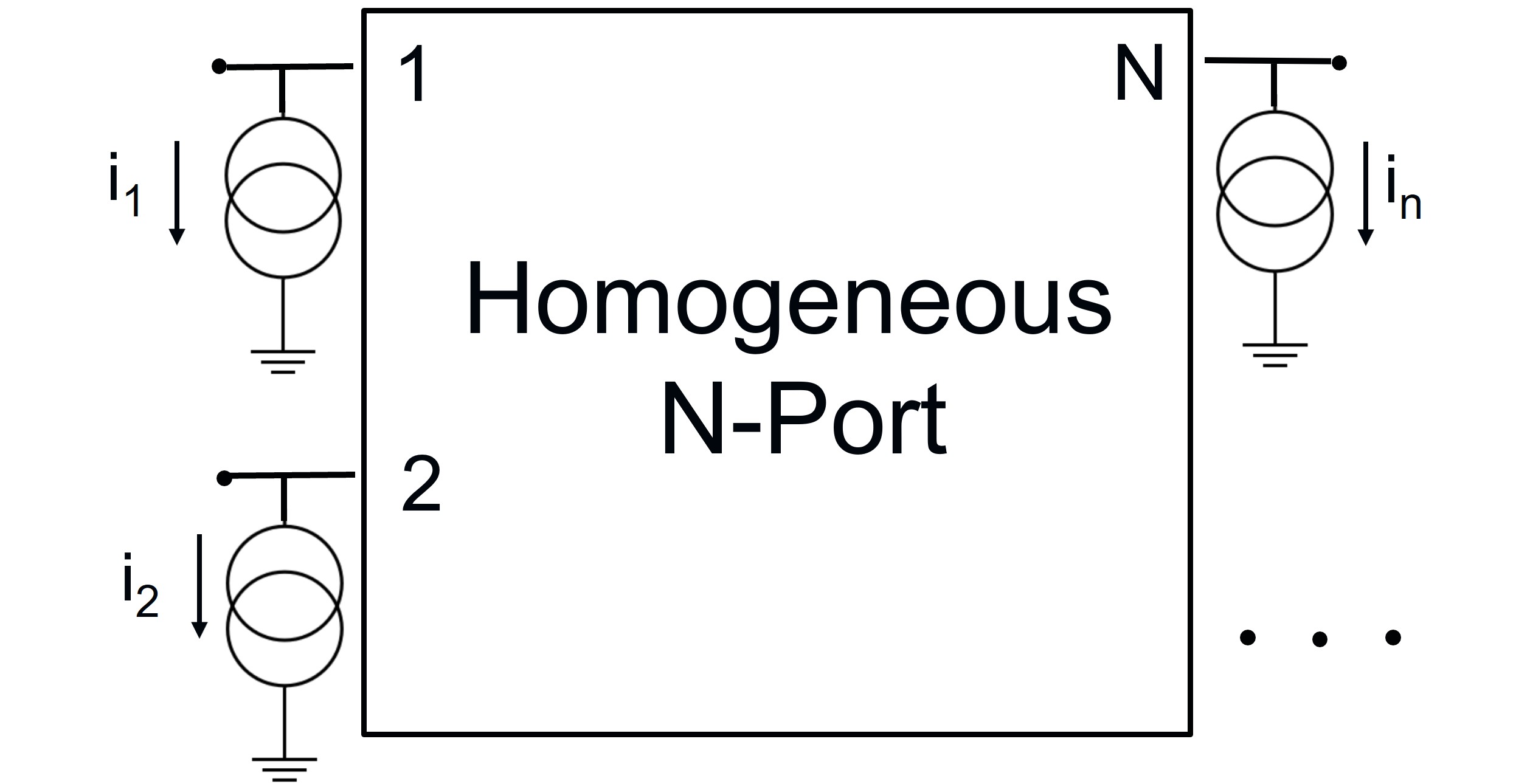}
\caption{Noisy $N$-Port Representation With Only Current Sources.}
\label{fig_three}
\end{figure}

When the admittance matrix $\mathbf{Y}$ is defined, where $\mathbf{Y}=\mathbf{Z}^{-1}$ when $\mathbf{Z}$ is non-singular and invertible,
we may similarly turn all voltage sources into current sources, so that
\begin{equation} 
\bar{\bar{\bar{\imath }}}_I=
\bar \imath-_{IJ}\bar v_J
\end{equation} 
and the equivalent representation in Fig.~\ref{fig_three} results. This representation is likewise 
completely equivalent to those in Figs.~\ref{fig_one} and \ref{fig_two}. 

We may also use an S-matrix travelling wave representation, where 
\begin{equation}
\begin{aligned}
v_I^{(in)}&=v^{(right)}_I=
v_I+Z_ci_I
\\
v_I^{(out)}&=v^{(left)}_I=
v_I-Z_ci_I
\end{aligned}
\end{equation}
and 
\begin{equation}
\mathbf{S}=
\frac{
\mathbf{Z}-Z_c\mathbf{I}
}{
\mathbf{Z}+Z_c\mathbf{I}
}.
\end{equation}
Here the characteristic impedance $Z_c$ is generally taken to be positive and real, corresponding
to some lossless transmission line, although a complex characteristic impedance is possible as well,
and the transmission line can be fictitious. 

\begin{figure}[!h]
\centering
\includegraphics[width=2in]{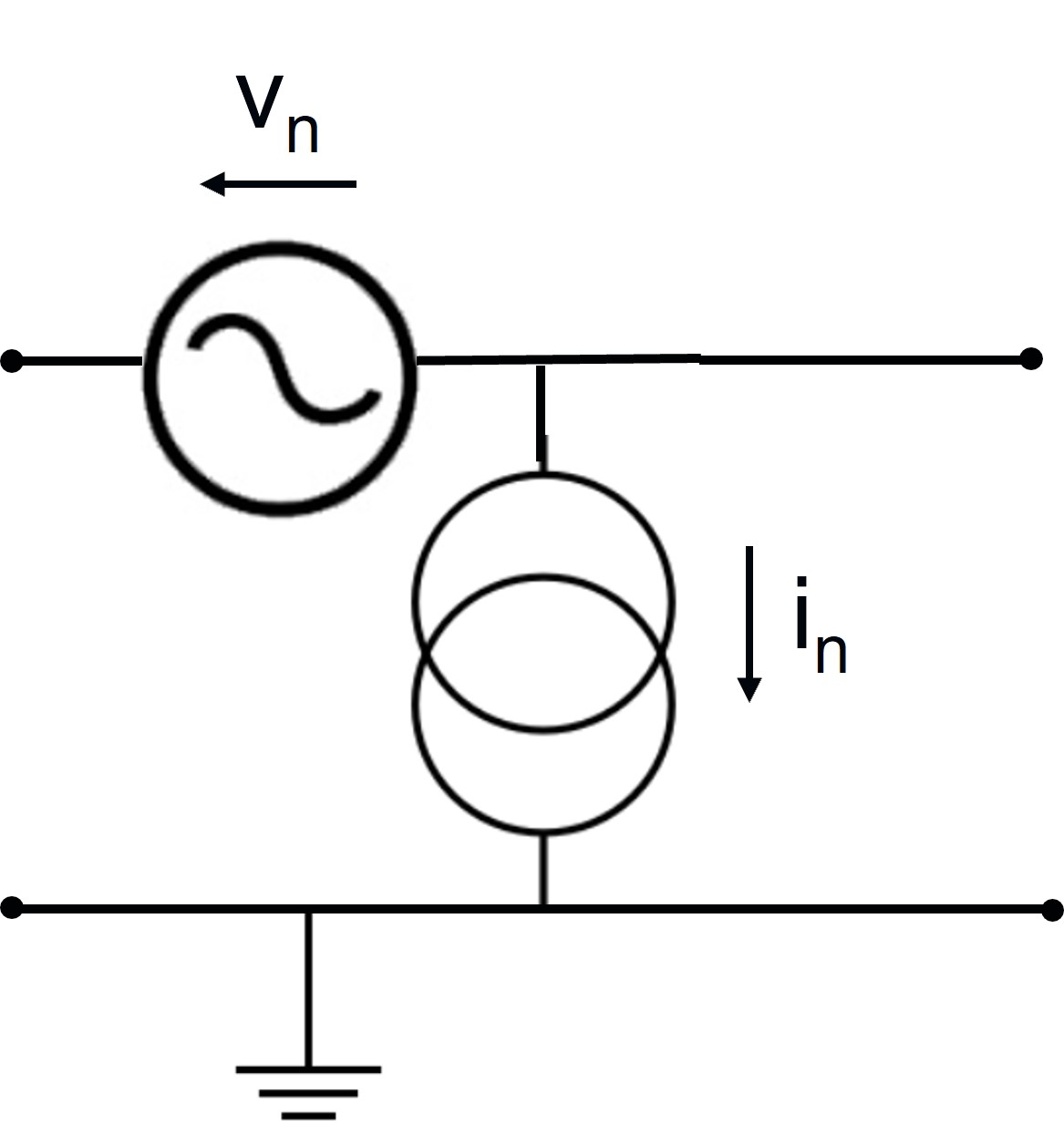} 
\includegraphics[width=2.5in]{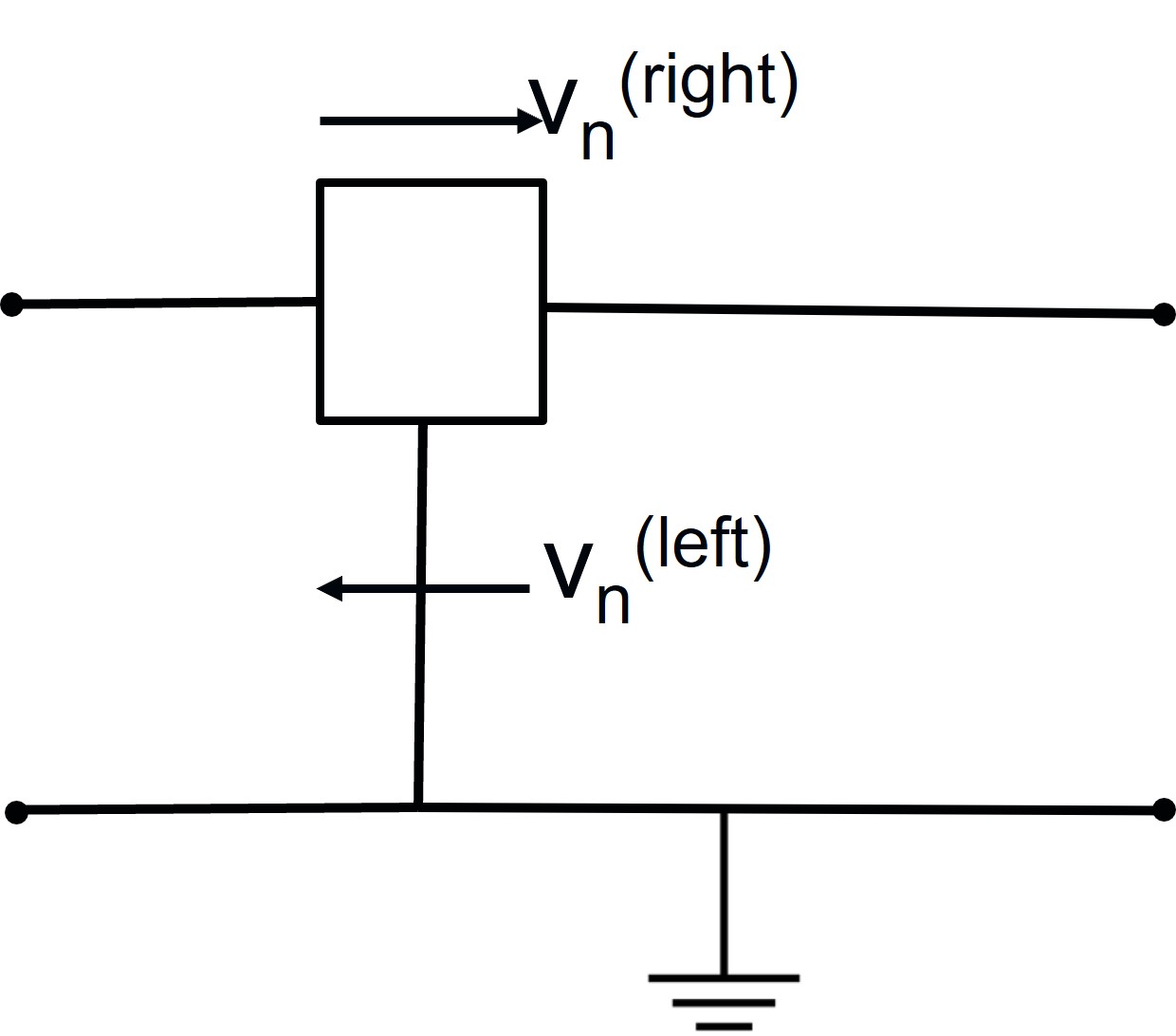} 
\caption{Travelling Wave Sources.}
\label{fig_five}
\end{figure}

In this travelling wave picture \cite{kurokawa,meys}, 
the voltage current source combination may be replaced by 
`in' and `out' travelling wave sources, as sketched in 
Fig.~\ref{fig_five}. Whereas in Fig.~\ref{fig_five}a the voltage and current
sources impose jump conditions between 
$v^{ext}$ and $v^{int},$ and 
between 
$i^{ext}$ and $i^{int},$ respectively, in Fig.~\ref{fig_five}b
the travelling wave sources impose jump conditions 
between $v^{in,ext}$ and $v^{in,int},$ and between $v^{out,ext}$ and $v^{out,int}.$

\section{Transfer Matrix Representation and its Generalizations}

Here we generalize the transfer matrix (or ABCD-matrix) representation
\begin{equation}
\begin{pmatrix}
v_2\\
i_2
\end{pmatrix}
=
\mathbf{T}_{21}
\begin{pmatrix}
v_1\\
i_1
\end{pmatrix}
=
\begin{pmatrix}
A_{21} &
B_{21}
\\
C_{21} &
D_{21}
\end{pmatrix}
\begin{pmatrix}
v_1\\
i_1
\end{pmatrix},
\end{equation}
which is so useful for analyzing a series of 2-ports chained together.
For the transfer matrix $\mathbf{T}_{21}$ the sign conventions
are typically modified a bit, so that a current following
into port 1 is considered positive whereas a current flowing
out of port 2 is regarded positive. For the inhomogeneous case, 
\begin{equation}
\begin{pmatrix}
v_2-\bar v_2
\\
i_2-\bar \imath _2
\end{pmatrix}
=
\mathbf{T}_{21}
\begin{pmatrix}
v_1-\bar v_1
\\
i_1-\bar \imath _1
\end{pmatrix},
\end{equation}
and we may find the alternative representation
\begin{equation}
\begin{pmatrix}
v_2-\hat v_2
\\
i_2-\hat \imath _2
\end{pmatrix}
=
\mathbf{T}_{21}
\begin{pmatrix}
v_1
\\
i_1
\end{pmatrix},
\end{equation}
where
\begin{equation}
\begin{pmatrix}
\hat v_2
\\
\hat \imath _2
\end{pmatrix}
=
\begin{pmatrix}
\bar v_2
\\
\bar \imath _2
\end{pmatrix}
-
\mathbf{T}_{21}
\begin{pmatrix}
\bar v_1
\\
\bar \imath _1
\end{pmatrix}.
\end{equation}
Here both the voltage and current sources on side 1 have been moved to
side 2. Since
$\mathbf{T}_{12}=
(\mathbf{T}_{21})^{-1},$
the opposite transformation is possible whenever
$\mathbf{T}_{12}$ is defined.

We begin by describing the plethora of equivalent representations of the
homogeneous $N$-port before describing the inhomogeneous case and then
the case where the inhomogeneous sources are described by a Gaussian
stochastic process, otherwise known as Gaussian noise.
While the 
$\mathbf{Z},$
$\mathbf{Y},$
and 
$\mathbf{S}$
matrix representations straightforwardly generalize to the $N$ port case, 
the mixed representations, of which the 2-port transfer matrix described above 
is an example, generalize into a large number of choices, which we now
describe in a perhaps less familiar, more mathematical manner, 
which facilitates transforming between the various representations, and 
moreover avoids having to deal separately with special singular cases.

Mathematically, a homogeneous $N$-port may be regarded as an 
$N$-dimensional linear subspace of the vector space 
$\mathbb{C}^{2N}$ whose coordinates may be taken to be
$v_1, \ldots, v_n, i_1, \ldots , i_N.$
Said another way, there are $N$ linearly independent linear equations,
or constraints, reducing 
$\mathbb{C}^{2N}$ to a subspace having the structure of
$\mathbb{C}^{N}.$ 
The choice of these $N$ equations, which may be represented as 
an $N\!\times \!2N$ rectangular matrix $\mathbf{M},$ is not unique:
one could have just a well chosen another set of equivalent equations
represented by 
$\mathbf{M}'= \mathbf{N} \mathbf{M}$ where
$\mathbf{N}$
is any invertible 
$N\!\times \!N$ square matrix. 

Generally, rather than the above representation, we want to partition 
the $2N$ variables into two sets such the values of the $N$ variables
of the first set are determined by the 
values of the $N$ variables of the second set
by means of a linear transformation.
These transformations will serve as a tool to eliminate sources on 
certain ports by moving them to ports on the left-hand side.

We start by counting the number of distinct ways to partition
$2N$ variables into two sets containing $N$ variables each where order
does not matter. There are
\begin{equation}
\begin{pmatrix}
2N \\ N~
\end{pmatrix}
=\frac{
(2N)!
}{
(N!)^2
}
\end{equation}
ways to partition the $2N$ variables into two sets of equal size, 
which we denote as 
$ L=( a_1, \ldots, a_N )$
and 
$ R=( b_1, \ldots, b_N ).$
The subspace $S$ defined by the $N$ linearly independent conditions may be 
expressed in block form, so that 
\begin{equation}
\mathbf{M}
\begin{pmatrix}
\mathbf{a}
\\ 
\mathbf{b}
\end{pmatrix}
=
(\mathbf{A}
~~\mathbf{B})
\begin{pmatrix}
\mathbf{a}
\\ 
\mathbf{b}
\end{pmatrix}
=0
\label{homoConstr}
\end{equation}
where 
$\mathbf{A}$ and 
$\mathbf{B}$
are $N\!\times \!N$ matrices
and eqn.~(\ref{homoConstr}) becomes
\begin{equation}
\mathbf{A}
\mathbf{a}
+
\mathbf{B}
\mathbf{b}
=0.
\end{equation}
When 
$\vert \mathbf{A}\vert \ne 0,$ 
\begin{equation}
\mathbf{a}
=-{\mathbf{A}}^{-1}
\mathbf{B}
\mathbf{b}.
\label{genRep}
\end{equation}
In this representation the $\mathbf{b}$ components 
are considered the independent variables and the 
$\mathbf{a}$ components the dependent variables.
When 
$\vert \mathbf{A}\vert =0,$ 
a singular case arises and no such representation is possible. 

We now generalize to the inhomogeneous case via the transformations
\begin{equation}
\begin{aligned}
v_I&\to ( v_I -\bar v_I), 
\\
i_I&\to ( i_I -\bar \imath _I)
\end{aligned}
\end{equation}
where $\bar v_I$ and $\bar \imath _I$ are the voltage and current sources
at port $I,$ respectively.

With the above substitutions, eqn.~(\ref{homoConstr}) may be generalized to become
\begin{equation}
\mathbf{M}
\begin{pmatrix}
v_1\\
\ldots \\
v_N \\ 
i_1\\
\ldots \\
i_N
\end{pmatrix}
=
\begin{pmatrix}
s_1\\
\ldots \\
s_N
\end{pmatrix}, 
\label{inhomoGenRep}
\end{equation}
and eqn.~(\ref{genRep}) becomes
\begin{equation}
\mathbf{a}=-
\mathbf{A}^{-1}
\mathbf{B}
\mathbf{b}
+
\mathbf{A}^{-1}
\mathbf{s},
\end{equation}
or 
\begin{equation}
(\mathbf{a}
-\hat {\mathbf{a}})
=-
\mathbf{A}^{-1}
\mathbf{B}
\mathbf{b}.
\end{equation}
We see that all the sources associated with the
$\mathbf{b}$ components have been eliminated in favor of sources
associated with only the
$\mathbf{a}$ components. 
Such a transformation possible whenever 
$\mathbf{A}$ is non-singular (i.e., an invertible
$N\!\times \!N$ matrix).

If we have $N$ sources with amplitudes
$\alpha _1, \ldots , \alpha _N,$ the Gaussian ensemble is described by the
probability distribution
\begin{equation}
p(\boldsymbol{\alpha })=
(2\pi )^{-N/2} \textrm{det}^{-1/2}(\mathbf{C})
\exp \left[
-\frac{1}{2}
\boldsymbol{\alpha }^\dagger
\mathbf{C}^{-1}
\boldsymbol{\alpha }
\right]
\end{equation}
where
\begin{equation}
C_{ij}=
\left\langle
{\alpha _i}^\dagger
{\alpha _j}
\right\rangle
\end{equation}
is the correlation matrix, which is Hermitian and positive definite
(i.e., all its eigenvalues are strictly positive).
The matrix $\mathbf{C}$ has $N^2$
independent real components.

\section{Measurement}

We shall assume that the homogeneous $N$-port parameters have already
been characterized, for example using a VNA, and are known. The 
conceptually most straightforward way of
characterizing the inhomogeneous parameters
would be for example one of the following: (i) each of the outputs
of the N-port could be shorted to ground with port $I$ terminal
connected to an ideal ammeter (of vanishing internal impedance), so
that $i_I$ at each port could be measured. Then
from theses measurement the correlation matrix 
$C^i_{IJ}(\omega )= \left\langle {i_I}^\dagger (\omega )~i_J(\omega )\right\rangle $
could be calculated. (ii) The same could be done with open circuit boundary
conditions at each port, with an idealized infinite internal impedance 
voltmeter attached between each port and ground so that $v_I(t)$ is 
monitored over a sufficiently long time, so that 
$C^v_{IJ}(\omega )= \left\langle {v_I}^\dagger (\omega )~v_J(\omega )\right\rangle $
can be evaluated. (iii) A hybrid method where some ports are shorted and others 
left open, with the current measured on the former and the voltage on the latter.
In each of these cases, the inhomogeneous $N$-port can be cast into the representation
eqn.~(\ref{inhomoGenRep}), which can be manipulated to any of the equivalent representations
described in the preceding sections. An alternative (iv) exists where  
each port is shunted to ground with a nonzero impedance $Z_I$ and the voltage at each port
is monitored. 

The above methods involve either measuring $N$ voltages and currents simultaneously, or 
equivalently making pairwise measurements between each pair of ports so that all the necessary
cross-correlations can be measured. But in some instances, for example in characterizing the 
noise properties of an amplifier, this would require measuring minute voltages or currents
at the input port. Much more accurate measurements can be obtained by placing different shunt
impedances across the input ports and measuring the total power at the output port. 

Let us for the moment assume that a voltages-only representation of the 
$N$-port is possible---in
other words, that the $\mathbf{Z}$-matrix description of the port is non-singular. (When this 
condition is not satisfied, the argument given here may be straightforwardly modified.) Let
us designate one of the ports, which we shall take to be the output port and label as port $N.$ We shall deduce the
the noise properties, more specifically the correlation function
$C^{v^*v}_{IJ}= \left\langle v^*_Iv_J^{\phantom{*}} \right\rangle ,$
based on total power measurements at port $N$
for a set of impedances at the ports $Z_I^{(a)},$ [$(a)=1\ldots \mathcal{N}$].
The data consists of the $\mathcal{N}$ real quantities $d^{(a)}=
\left\langle v^*_Nv_N^{\phantom{*}} \right\rangle (Z_1^{(a)}, \ldots , Z_N^{(a)}).$
No cross correlations are measured directly. Rather these are inferred from the impedance
dependence of the total power measured at port $N.$ 
The correlation matrix $C^{v^*v}_{IJ}$
has $N^2$ independent real parameters, 
so one would naively expect 
$C^{v^*v}_{IJ}$ to be determined whenever 
$\mathcal{N}\ge N^2$ assuming that none of the linear algebra is singular.
Mathematically, port $N$ could be any port, but as a practical matter we would
want to chose the measurement port for maximal noise output power, in order to maximize
the accuracy of the measurement and the minimize the requirements on the measurement
apparatus. Once $\left\langle v^*_Iv_J^{\phantom{*}} \right\rangle $
has been measured, as was described above, 
the noise properties can be transformed to any other representation
for which the transformation is non-singular. 

\section{Conclusion}

We have explored the large number of equivalent representations
of an inhomogeneous $N$-port and the transformations relating them.
In a nutshell, when there are no obstructions arising from singularities
in the linear algebra, it suffices to place $N$ voltage or current sources
at the output ports of the embedded homogeneous $N$-port
so that each port has at most one voltage source and one current source.
An alternative description in terms of travelling wave sources is also described.
The most general Gaussian noise of a noisy $N$-port
is described by an $N\times N$ complex Hermitian correlation matrix for these complex sources, corresponding to $N^2$ real degrees of freedom.
When a number of such multi-ports are connected together to form a new port, it is possible by means of the transformations described in this paper to move 
the noise sources so that they are all situated on the external legs of the 
new port---that is, so that one has an embedded homogeneous port with noises source as 
described above. We also described how to measure this correlation matrix.

\section*{Acknowledgments}

\noindent
MB acknowledges a SKA-LOFAR travel grant from the Observatoire de Paris
for a trip to Cambridge where part of this work was done. 
MB and DM thank Eloy de Lera Acedo, Christophe Craeye, and Dirk de Villiers for useful discussions and comments. DM also acknowledges Murray Edwards College of the University of Cambridge.


\begin{IEEEbiography}%
[{\includegraphics[width=1in,height=1.25in,clip,keepaspectratio]%
{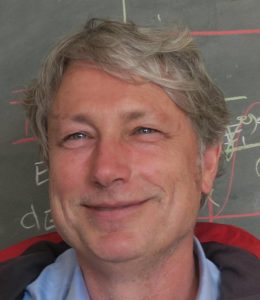}}]
{Martin Bucher}
is with the CNRS based at the
Laboratoire APC (Astroparticules et Cosmologie) at
Universit\'e Paris Cit\'e in Paris, France. Bucher
received his PhD in Physics from Caltech in 1990
and has held positions at the Institute of Advanced Study,
Princeton University, Stony Brook University, and DAMTP
at the University of Cambridge. Bucher held the SW Hawking
Fellowship of Mathematical Sciences at Trinity Hall, University
of Cambridge from 2000-2004 before coming to France where he
worked on the ESA Planck Mission, which mapped the microwave sky
in nine frequencies, thus constraining models of the primordial
universe. Bucher was recipient of the 2018 Gruber Prize 
in Cosmology
as part of the Planck team and is member of the Academy of 
Science of South Africa. Bucher is a part of the REACH 
collaboration. 
\end{IEEEbiography}

\begin{IEEEbiography}%
[{\includegraphics[width=1in,height=1.25in,keepaspectratio]
{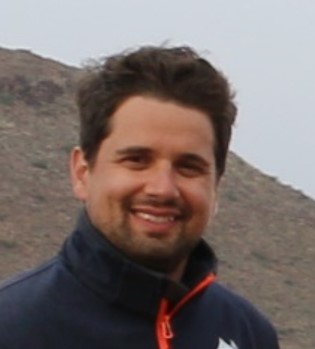}}]{Daniel Molnar}
is with the University of Cambridge. 
Daniel Molnar received an MPhil degree and
Ph.D.~in Physics from the University of
Cambridge, Cambridge, UK, in 2013 and 2016.
He has worked at the Department of Engineering
at the University of Cambridge, COMSOL Inc., and
the Theory of Condensed Matter group at the
Cavendish Laboratory, University of 
Cambridge. He later worked at CERN, the European
Particle Accelerator Laboratory, in Geneva, 
Switzerland as a Senior Fellow.
Dr. Molnar is a Bye-Fellow in Physics at Murray
Edwards College at the University of Cambridge.
Molnar is a member of the REACH Collaboration.
\end{IEEEbiography}

\vfill

\end{document}